\documentclass[numberedappendix,apj,twocolumn,letter]{emulateapj}

\usepackage{amssymb}
\usepackage{amsmath}
\newcommand{\be}{\begin{equation}}
\newcommand{\ee}{\end{equation}}

\shorttitle{Galaxy Clustering at $z\sim8$ from the BoRG Survey}
\shortauthors{Trenti et al.}

\begin{document}


\title{Overdensities of Y-dropout Galaxies from the Brightest-of-Reionizing Galaxies Survey: A Candidate Protocluster at Redshift
  $z\approx 8$\footnote{Based on observations made with the NASA/ESA
    Hubble Space Telescope, which is operated by the Association of
    Universities for Research in Astronomy, Inc., under NASA contract
    NAS 5-26555. These observations are associated with Program
    $11700$.}}

\author{Michele Trenti} \affil{University of Colorado, Center for Astrophysics and Space Astronomy, 389-UCB, Boulder, CO 80309 USA} \email{trenti@colorado.edu} 
\and
\author{L.~D. Bradley}\affil{Space Telescope Science Institute, 3700 San Martin Drive Baltimore MD 21218 USA}
\and \author{M. Stiavelli} \affil{Space Telescope Science Institute, 3700 San Martin Drive Baltimore MD 21218 USA}
\and \author{J.~M. Shull}\affil{CASA, Department of Astrophysical and Planetary Science, University of Colorado, 389-UCB, Boulder, CO 80309 USA}
\and \author{P. Oesch} \affil{Astronomy \& Astrophysics Department, University of
  California, Santa Cruz, CA 95064, USA; Hubble Fellow}
\and \author{R.~J. Bouwens} \affil{Astronomy \& Astrophysics Department, University of California, Santa Cruz, CA 95064, USA ; Leiden Observatory, University of Leiden, Postbus 9513, 2300 RA Leiden, Netherlands}
\and \author{J.~A. {Mu{\~n}oz}}\affil{University of California Los Angeles, Department of Physics
and Astronomy; Los Angeles, CA 90095, USA}
\and \author{E. Romano-Diaz}\affil{Department of Physics and Astronomy,
University of Kentucky, Lexington, KY 40506-0055, USA; Argelander Institut fuer Astronomie, Auf dem Haegel 71, D-53121 Bonn, Germany}
\and \author{T.~Treu}\affil{Department of Physics, University of California, Santa Barbara, CA 93106-9530, USA}
\and \author{I. Shlosman} \affil{Department of Physics and Astronomy,
University of Kentucky, Lexington, KY 40506-0055, USA}
\and \author{C.~M. Carollo} \affil{Institute of Astronomy, ETH Zurich, CH-8093 Zurich, Switzerland}


\begin{abstract}
  Theoretical and numerical modeling of the assembly of dark-matter
  halos predicts that the most massive and luminous galaxies at high
  redshift are surrounded by overdensities of fainter companions. We
  test this prediction with {\it{Hubble Space Telescope}} observations
  acquired by our Brightest of Reionizing Galaxies (BoRG) survey,
  which identified four very bright $z\sim 8$ candidates as
  $Y_{098}$-dropout sources in four of the 23 non-contiguous WFC3
  fields observed. We extend here the search for $Y_{098}$-dropouts to
  fainter luminosities ($M_*$ galaxies with $M_{AB}\sim -20$), with
  detections at $\geqslant 5\sigma$ confidence (compared to the
  $8\sigma$ confidence threshold adopted earlier) identifying 17 new
  candidates.  We demonstrate that there is a correlation between
  number counts of faint and bright $Y_{098}$-dropouts at $\geqslant
  99.84\%$ confidence. Field BoRG58, which contains the best bright
  $z\sim 8$ candidate ($M_{AB}=-21.3$), has the most significant
  overdensity of faint $Y_{098}$-dropouts. Four new sources are
  located within $70''$ (corresponding to $3.1$ comoving
  $\mathrm{Mpc}$ at $z=8$) from the previously known brighter $z\sim
  8$ candidate. The overdensity of $Y_{098}$-dropouts in this field
  has a physical origin to very high confidence ($p>99.975\%$),
  independent of completeness and contamination rate of the
  $Y_{098}$-dropout selection. We modeled the overdensity by means of
  cosmological simulations and estimate that the principal dark matter
  halo has mass $M_h\approx(4-7)\times 10^{11}~\mathrm{M_{\sun}}$
  ($\sim 5\sigma$ density peak) and is surrounded by several
  $M_h\approx 10^{11}~\mathrm{M_{\sun}}$ halos which could host the
  fainter dropouts. In this scenario, we predict that all halos will
  eventually merge into a $M_h>2\times10^{14}~\mathrm{M_{\sun}}$
  galaxy cluster by $z=0$.  Follow-up observations with ground and
  space based telescopes are required to secure the $z\sim 8$ nature
  of the overdensity, discover new members, and measure their precise
  redshift.

\end{abstract}

\keywords{galaxies: evolution --- galaxies: high-redshift}

\section{Introduction}

A growing number of galaxies at $z>7$ is being discovered with
Wide Field Camera 3 (WFC3) on the Hubble Space Telescope
\citep{oesch10,bouwens10c,mclure10,finkelstein10,wilkins10,yan11,trenti11}. These
galaxies are identified from deep imaging in multiple bands, showing
the signature of absorption of photons with wavelength shorter than
Lyman-$\alpha$ (Ly$\alpha$) at $121.6~\mathrm{nm}$ by intervening
neutral hydrogen \citep{steidel96}. Based on the wavelength range of
the filter in which the photometric break is observed, an approximate
redshift is derived. The Lyman-break technique has been validated
extensively with spectroscopic follow-ups of large samples at
$z\lesssim7$
\citep{malhotra05,stark10,ono11,pentericci11}. Non-ambiguous redshift
measurements for galaxies at $z>7.2$ are missing despite several
attempts, possibly because of an increase in the neutral fraction of
hydrogen compared to lower redshift \citep{lehnert10,schenker11}.

HST observations and their ground-based spectroscopic follow-ups are
providing a glimpse of the formation of stars and galaxies during the
epoch of hydrogen reionization at $z>7$, showing a declining star
formation rate as the redshift increases and a rapidly evolving galaxy
luminosity function \citep{bouwens11}. The evolution of the luminosity
function is consistent with the expected evolution of the dark-matter
halo mass function: at higher redshift, massive dark-matter halos
become progressively rarer and the typical mass evolves toward smaller
values \citep{trenti10,finlator10,jaacks11}. It is thus not surprising that the
majority of $z>7$ galaxies, identified from pencil-beam surveys such
as the HUDF, are faint, with magnitudes $m_{J125}>27$.

To search for the brightest and rarest $z\sim8$ galaxies, we are
carrying out a large area medium-deep HST survey along random lines of
sight (\citealt{trenti11}, hereafter T11). Bright galaxies, such as
those identified by our Brightest of Reionizing Galaxies (BoRG)
survey, are expected to be highly clustered and to be found
preferentially in groups rather than in isolation. This scenario has
been verified observationally out to $z\lesssim 6.5$ by several
groups, both in the context of Ly$\alpha$ emitters
\citep{shimasaku03,venemans04,ouchi05} and of Lyman-break galaxies
\citep{overzier06a,zheng06,cooke08,utsumi10,capak11}. Clustering of sources
around a rare, bright galaxy extends to less luminous companions,
under a broad assumption that galaxy luminosity is correlated with the
mass of the dark-matter host halo. This provides an indirect test to
investigate whether the bright galaxy is at high redshift rather than
a foreground contaminant that entered into the Lyman-break photometric
selection window \citep{munoz08}.

In T11, we identified four $z\sim 8$ candidates with $25.5\leq
m_{J125} \leq 26.7$ and detections at $\geqslant 8\sigma$ confidence
from an area of approximately $130$ arcmin$^2$.  In this paper we
revisit the search for $z\sim 8$ candidates in the BoRG survey,
extending the selection to fainter sources, detected at $\geqslant
5\sigma$ confidence. Our goal is to test whether the brightest $z\sim
8$ candidates are preferentially associated with overdensities of
fainter dropouts, thereby providing support for the $z\sim 8$ nature
of the bright candidate based on the findings by \citet{munoz08}. Our
particular focus is on field BoRG58, which contains the $z\sim8$
galaxy with the highest signal-to-noise ratio (S/N) found in the
survey so far, with J-band magnitude $m_{J125}=25.8\pm0.2$ (T11). The
photometric quality of the data available for the field allow us to
extend the search for fainter dropout sources with
$m_{J125}\lesssim27.0$.

This paper is organized as follows. In Section~\ref{sec:survey} we
summarize the data reduction and analysis, and in
Section~\ref{sec:clustering_survey} we investigate the clustering of
the newly identified $Y_{098}$-dropouts. In Section~\ref{sec:tests} we
discuss the properties of the $Y_{098}$-dropouts in field BoRG58 and
perform extensive tests to evaluate the likelihood that we are
observing a 3D, physical overdensity in that
field. Section~\ref{sec:theory} compares our observations to
theoretical/numerical models. Section~\ref{sec:conclusion} summarizes
our findings and concludes by discussing the outlook for follow-up
observations to strengthen the results obtained in this work. We adopt
WMAP5 cosmology \citep{wmap5} and the AB magnitude scale \citep{oke}.

\section{Data Reduction and Dropout Selection}\label{sec:survey}

The BoRG survey is a four-band (F606W, F098M, F125W, F160W)
pure-parallel HST/WFC3 survey designed primarily to identify bright
galaxies ($m_{J125}\lesssim27$) at $z\gtrsim7.5$ based on their
broad-band colors using the Lyman-break technique
\citep{steidel96}. Data are associated primarily with program GO-11700
(PI Trenti), complemented by program GO-11702 (PI Yan; see
\citealt{yan11}). In this paper we consider the data from GO-11700 (23
out of 29 WFC3 pure-parallel pointings) to guarantee a uniform design
of the observations, as described in T11. Exposure times vary from
pointing to pointing because of the pure-parallel nature of the survey
and are reported, along with the $5\sigma$ limiting magnitudes in
Table 1 of T11.  For reference, we report here the exposure times and
limits for field BoRG58, which contains the best $z\sim8$ candidate
identified in T11.  That field, centered at
$\mathrm{RA}=219^{\circ}.230$,
$\mathrm{Decl}=+50^{\circ}.719$ (J2000 system), was observed with
the following exposure times (also indicated is the $5\sigma$
sensitivity within an aperture of radius $r=0''.32$).  F125W:
$t=2509~\mathrm{s}$ ($m_{lim}=27.0$); F160W: $t=1806~\mathrm{s}$
($m_{lim}=26.6$); F098M: $t=4912~\mathrm{s}$ ($m_{lim}=27.1$); F606W:
$t=2754~\mathrm{s}$ ($m_{lim}=26.8$).
 
The observations are designed so that at least two different exposures
per filter were taken in each field, but there is no dithering, as
the spectroscopic primary observations do not use it. Individual
exposures have been calibrated first with calwfc3, then aligned and
registered on a common $0.08~\mathrm{arcsec/pixel}$ scale using
\texttt{multidrizzle} \citep{koekemoer2002}.

We identified sources from the $J_{125}$ images using SExtractor in
dual image mode \citep{bertin96}, requiring at least 9 contiguous
pixels with $S/N\geq0.7$ after normalization of the RMS maps as
discussed in T11. Sources are accepted in our final catalog if they
pass a minimum signal-to-noise (S/N) threshold with $S/N\geq5$ in
$J_{125}$ and $S/N\geq2.5$ in $H_{160}$ for ISOMAG fluxes. For
Gaussian noise, uncorrelated in the two bands, there is a probability
$p<6.5\times10^{-9}$ that a noise fluctuation satisfies our
source-selection criteria. This illustrates that the likelihood of
introducing spurious sources within this catalog is very low, as
demonstrated quantitatively in Section~\ref{sec:tests}.

\subsection{Lyman-break selection and contamination}\label{sec:lbg_selection}

To select $z\sim8$ candidates, we impose the following conditions on
the sources in the catalog:

\begin{equation}
m_{Y098}-m_{J125}\geq1.75;
\end{equation}
\begin{equation}\label{eq:J-H}
m_{J125}-m_{H160}<0.02+0.15\times(m_{Y098}-m_{J125}-1.75),
\end{equation}
and we finally require non-detection in $V_{606}$ ($S/N<2$). Flux
measurements are corrected for foreground Galactic extinction using
the maps by \citet{schlegel98}. For the determination of the
$m_{Y098}-m_{J125}$ color, we replace the flux in $Y_{098}$ with its
$1\sigma$ upper limit if the measurement is below that limit. This is
the standard practice for the Lyman-break galaxy selection (e.g.,
\citealt{bouwens07}). A different selection technique, based on the
fit of the observed fluxes in all bands available to construct a
photometric redshift probability distribution, may resort to $2\sigma$
upper limits \citep[e.g., see][]{mclure10}. In this respect, we note
that our $1\sigma$ flux-replacement for the Y-J color would be
equivalent to a $2\sigma$ replacement while at the same time requiring
a slightly less prominent break, that is
$m_{Y098}-m_{J125}\geq1.5$. Such selection is still more conservative
than $m_{Y098}-m_{J125}\geq1.25$ adopted by \citet{bouwens10c}, which
nevertheless resorts to $1\sigma$ limits.

We measure the near-IR colors based on isophotal photometry without
Point Spread Function (PSF) corrections, an approach that optimizes
the signal-to-noise ratio, as determined by extensive Monte-Carlo
simulations of artificial source recovery \citep{oesch07}. This
approach has been adopted in several other studies
(\citealt{oesch09,oesch10}; T11). Alternative approaches include
resorting to either fixed apertures (see \citealt{bunker10}) or to
Kron-line elliptical apertures (see \citealt{bouwens07}). The variety
of approaches present in the literature suggests that a gold standard
is missing, primarily because results are comparable among all
different methods, as shown by \citet{finkelstein10}. In addition, we
verified directly that the impact of PSF corrections on the colors of
the $Y_{098}$-dropouts in the BoRG survey is small. For this purpose,
we corrected the isophotal photometry based on the WFC3 PSF model
provided by
STScI\footnote{\url{http://www.stsci.edu/hst/wfc3/ir\_ee\_model\_smov.dat}},
interpolating the data provided over wavelength and radius. For the
radius, we defined an effective aperture radius for each dropout based
on the isophotal area used for the photometry. Faint dropouts are
typically detected over an average of 20 pixels, that is an effective
radius of $r_{\mathrm{eff}}\sim 0''.2$. This leads to small
corrections (compared to the random error on the colors):
$\Delta_{Y098-J125}=0.11$ and $\Delta_{J125-H160}=0.14$.

\begin{figure} 
\includegraphics[scale=0.44]{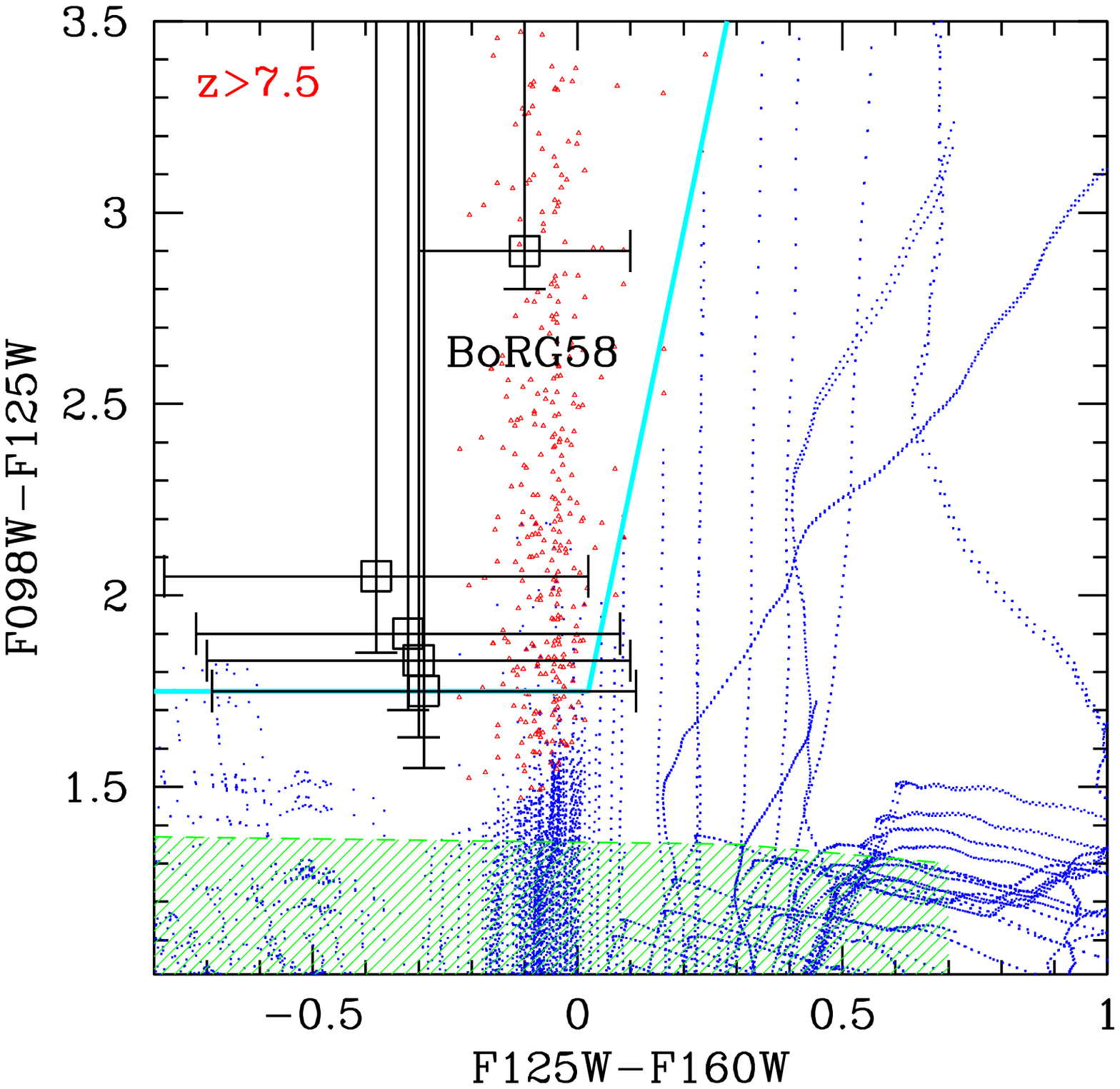}  
\includegraphics[scale=0.38]{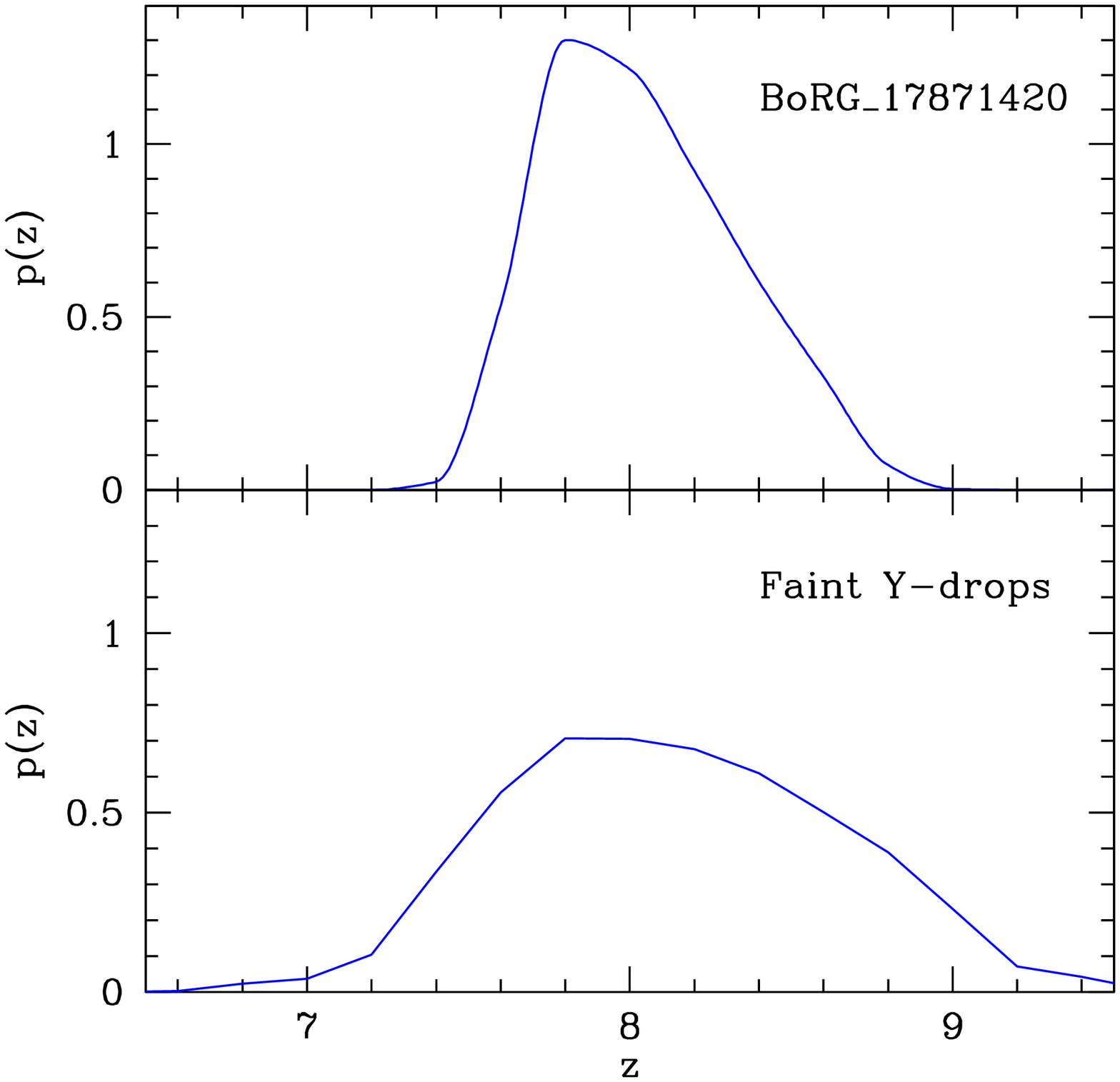}
\caption{Left panel: F098M-dropouts color-color selection. Black
 squares indicate the five $z\approx8$ candidates (with $1\sigma$
 error-bars on colors). Cyan lines denote the selection window. Blue
 dots are simulated low-redshift interlopers; red triangles are
 $z>7.5$ galaxies. L/T dwarf stars would appear in green
 shaded area. Right panel: photometric redshift distribution
 derived from Monte-Carlo source-recovery simulations (top:
 BoRG58\_17871420; bottom: fainter sources).}\label{fig:colcol}
\end{figure}

Our color-color selection includes objects in the redshift range
$7.5\lesssim z\lesssim8.5$ and is efficient at rejecting low-redshift
interlopers and brown-dwarf stars (Figure~\ref{fig:colcol}; see also
\citealt{bouwens10c}). The most likely contaminants in the color
selection are $1.5\lesssim z\lesssim2$ galaxies with either a
prominent Balmer break or with strong emission lines. In T11, we
estimated a contamination fraction $\lesssim30\%$ based on population
synthesis modeling, which includes templates with emission lines, and
comparison with the GOODS/ERS data for the bright dropouts discussed
there.

Recently, \citet{atek11} extrapolated data from a spectroscopic survey
shallower than BoRG to derive a contamination rate from low-$z$
galaxies with strong emission lines in the BoRG survey of one bright
$Y_{098}$-dropout every 17 fields. They suggested that galaxies with a
very faint continuum and flux in the J and H band given primarily by
line emission (\ion{O}{2}, $\lambda=3727 \mathrm{\AA}$; \ion{O}{3},
$\lambda=5007~ \mathrm{\AA}$, and $H\alpha$, $\lambda = 6563~
\mathrm{\AA}$) can enter our photometric selection window. For a
galaxy with $m_{J125}=26$, this implies a line flux $\phi\sim
8\times10^{-17}~\mathrm{erg~cm^{-2}~s^{-1}}$. Observationally, we have
no evidence that this is the case for the bright $Y_{098}$-dropout in
field BoRG58, which has been followed-up with Keck spectroscopy.
\citet{schenker11} observed that bright $z\sim 8$ galaxy candidate
with NIRSPEC, covering the wavelength range $0.95~\mathrm{\mu m} \leq
\lambda \leq 1.29~\mathrm{\mu m}$ and reaching a median $10\sigma$
sensitivity of $2.2 \times 10^{-17}~\mathrm{erg~cm^{-2}~s^{-1}}$ in the
N1 filter and $4.3\times 10^{-17}~\mathrm{erg~cm^{-2}~s^{-1}}$ in the
N2 filter. No emission lines were detected. If the candidate had been
an emission-line contaminant as suggested by \citet{atek11}, the line
would have been clearly detected (at $\gtrsim 20 \sigma$ confidence),
unless the contaminant has an \ion{O}{3} line located in a very narrow
region of the spectrum with $\Delta \lambda \sim 0.01~\mathrm{\mu m}$
around $\lambda = 1.30~\mathrm{\mu m}$. All other configurations are
ruled out. In fact, if the flux in J is due to \ion{O}{3}
($2.0<z<2.7$), \ion{O}{3} moves redward of the $H_{160}$ filter for
\ion{O}{2} at $\lambda \geq 1.27~\mathrm{\mu m}$. If instead the
$J_{125}$ flux is due to \ion{O}{3} ($1.2<z<1.8$), then $H\alpha$
moves out of $H_{160}$ for \ion{O}{3} at $\lambda \geq
1.30~\mathrm{\mu m}$. Spectroscopic follow-up is therefore efficient
at identifying this class of contaminants. In the case of the BoRG58
field, the observations of \citet{schenker11} essentially rule out
that the brightest $Y_{098}$-dropout is an emission-line contaminant.

Because we are focusing here on fainter sources, we revisited our
estimate of contamination. For the fainter dropouts near the selection
limit of our survey, we estimate that contamination can reach $\sim
50\%$. This is comparable to that reported by \citet{bouwens10c} for
ERS $Y_{098}$-dropouts (where there is an expected $\sim 38\%$ contamination;
see also \citealt{mclure11,capak11b}). Our data are less deep in the
optical bands, but we impose more stringent near-IR color-color cuts,
which reduce the impact of photometric scatter at the expense of
reduced completeness of our faint sample.

Other sources of possible contaminants are galactic ultra-cool stars,
which have red $Y-J$ colors \citep{knapp04}. However, the use of F098M
as a dropout filter is efficient at rejecting this class of contaminants
when combined with our selection criteria $Y_{098}-J_{125}>1.7$. As
shown in Fig.~\ref{fig:colcol}, the locus of brown dwarfs is well
separated from our selection window for
$Y_{098}$-dropouts. \citet{ryan11} searched for brown dwarf stars in a
data-set that included all the BoRG fields considered here and
identified a total of 17 ultra-cool stars in $230$ arcmin$^2$. All
their candidates have $Y_{098}-J_{125}<1.4$, so it is unlikely that
any faint and unresolved source with $Y_{098}-J_{125}>1.7$ has stellar
origin.

\section{Correlation between faint and bright $Y_{098}$-dropouts in the BoRG survey}\label{sec:clustering_survey}

The $z\sim8$ Lyman-break selection for faint sources discussed in
Section~\ref{sec:lbg_selection} has been applied to the 23 BoRG fields
from program GO-11700, with the goal of investigating whether the BoRG
survey contains evidence of clustering of fainter $Y_{098}$-dropouts around
the brighter sources identified in T11. Sources have been inspected
visually to ensure that they are not artifacts such as hot pixels,
diffraction spikes of stars, or outer regions of brighter foreground
galaxies. 

\begin{figure} 
\includegraphics[scale=0.8]{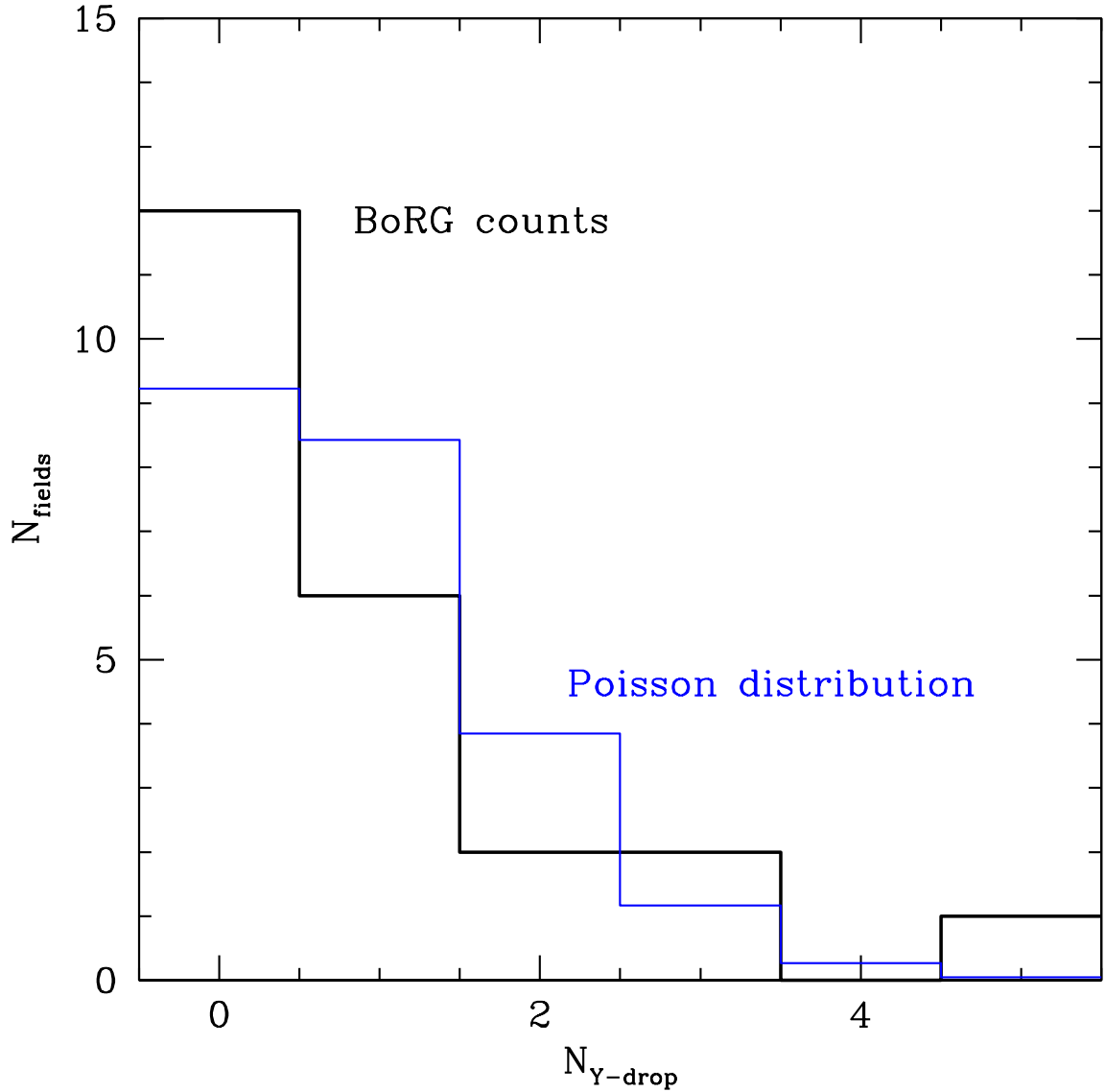}
\caption{Number counts distribution of the Y-band dropouts within the
  23 BoRG fields considered in this paper (solid histogram). Blue
  lines show a Poisson distribution with the same mean ($\langle N
  \rangle = 0.913$ per field).}\label{fig:number_counts}
\end{figure}

A total of 21 sources satisfy the $Y_{098}$-dropout selection
(Table~\ref{tab:field_summary}). Therefore, we identify 17 new
candidates (defined as ``faint'' sample) in addition to the four
published in T11 (defined as ``bright'' sample). The ``bright'' versus
``faint'' classification is based on a S/N selection in the $J_{125}$
band, with bright meaning $S/N \geq 8$. The number counts per field of
the 21 candidates are distributed as follows. Twelve fields have no
sources satisfying our selection criteria, and six fields contain a
single faint candidate. The remaining five fields have multiple
candidates. Four of these five fields are those that contain one of
the bright candidates published in T11.  In BoRG1k and BoRG66 we find
an additional $Y$-dropout. BoRG0t has two faint dropouts in addition
to the bright one we already reported, while four additional
candidates have been discovered in field BoRG58. Finally, three
candidates are present in field BoRG0y, including one with
$m_{J{125}}=26.6$ that barely misses the S/N cutoff adopted in T11
($J_{125}$ detection at $7.9\sigma$ confidence, $H_{160}$ at
$4.5\sigma$). Table~\ref{tab:field_summary} summarizes our findings.
The distribution of the number counts is presented in
Figure~\ref{fig:number_counts} and compared to the expectation from
the null hypothesis (no clustering), which is a Poisson distribution
with average value $\langle N \rangle = 0.913$. This comparison has
been suggested as a method to infer the clustering properties of
dropout samples in pure-parallel surveys \citet{robertson10}. Visual
inspection immediately identifies the excess at $N=5$ (field BoRG58),
which will be investigated in detail in the subsequent sections of the
paper. The remaining of the number-counts distribution appears to
deviate slightly from Poisson, but because of the small number of
fields and dropouts per field, a marked difference is neither expected
nor statistically significant. At low number counts, Poisson
uncertainty dominates the contribution to the standard deviation of
the distribution \citep{ts08,robertson10}.

To analyze the clustering of dropouts, there is however fundamental
information that is not conveyed in Figure~\ref{fig:number_counts}:
All four fields with a bright $z\sim8$ candidate from T11 contain at
least a second, fainter, $Y_{098}$-dropout. To investigate the chance
of this happening, we resorted to a Monte-Carlo simulation to derive
the probability that a random distribution of the 17 new sources among
the 23 fields analyzed leads to eight or more faint sources in the
four fields with a brighter candidate, with a minimum of one source
for each field. We carried out $10^5$ random realizations and derived
that such occurrence has probability $p=0.16\%$.  Therefore, the
association between the newly discovered dropouts and the brighter
previously reported sources is significant at greater than $99.84 \%$
confidence. The presence of multiple dropouts in field BoRG0y
qualitatively strengthens the confidence of the result, as the
brightest dropout in that field barely misses the S/N cutoff of T11
and therefore our definition of ``bright'' dropout.

Our choice to define ``bright'' and ``faint'' dropouts based on a S/N
measurement rather than absolute luminosity minimizes the impact of
differences in the exposure times among the BoRG fields. In fact, the
luminosity function of $Y_{098}$-dropouts around the detection limit
of the BoRG survey, $\langle m_{lim} \rangle =26.7$, can be locally
approximated by a power law, so that the relative expected abundance
of bright and faint sources is approximately uniform across the
survey. In addition, we verified that the correlation in the number
counts between the faint and bright sample of dropouts is not arising
because the fields containing bright candidates have deeper
integration times. The integration time of the fields containing a
dropout identified in T11 is representative for the survey: of the
four, two have below median depths and two are above
median. In particular, field BoRG58, containing the most significant
excess of dropouts, has an exposure time $t=2500$ s in F125W, only
slightly deeper than the median exposure of the survey ($t=2200$ s in
F125W).

Foreground Galactic reddening is not the cause of the observed
correlation between faint and bright dropouts. From the
\citet{schlegel98} maps, all fields with $z\sim 8$ candidates have low
extinction values ($A_B\leq 0.24$). The average extinction for the
sample of the four fields with bright candidates is $\langle A_B
\rangle =0.07 \pm 0.04$ (and BoRG58 has $A_B=0.06$), while the average
extinction for the fields containing only faint candidates is $\langle
A_B\rangle =0.17 \pm 0.07$. Such low extinction values imply that the
reddening corrections we apply to our near-IR photometry are very
small ($\Delta m \lesssim 0.05$).

Finally, we note that the analysis presented here on the correlation
between bright and faint dropouts is completely independent of
contamination and completeness of the selection, as it is uniquely
based on number counts. Both contamination and incompleteness actually
contribute toward reducing the significance of clustering. In fact, if
the sample of dropouts consists of a population of bona-fide $z\sim 8$
objects plus some contaminants at $1\lesssim z \lesssim 2$, no
angular cross-correlation is present between the two populations. In
such case, the clustering signal is less strong than for the case in
which the whole sample is in the same redshift
interval. Incompleteness also reduces clustering. For a fixed
number of objects, an incomplete sample contains objects at fainter
luminosities compared to a complete sample. Because galaxy luminosity
correlates with halo mass, this reduces the average bias of
the population \citep{ts08}\footnote{This effect is also illustrated
  by our public cosmic variance calculator available at
  \url{http://casa.colorado.edu/~trenti/CosmicVariance.html}}.

So far, our analysis has focused only on the global correlation of
counts in the survey, showing that faint dropouts are preferentially
found in fields with bright dropouts at a confidence level greater
than 99.84\% {\emph{regardless of contamination and completeness of
    the selection}}. In the next Section we investigate in details the
properties of field BoRG58, which contains the highest number of
dropouts among the BoRG survey fields.

\section{Candidate $z\sim 8$ Lyman-Break Galaxies in Field
  BoRG58}\label{sec:tests}

\begin{figure} 
\includegraphics[scale=0.41]{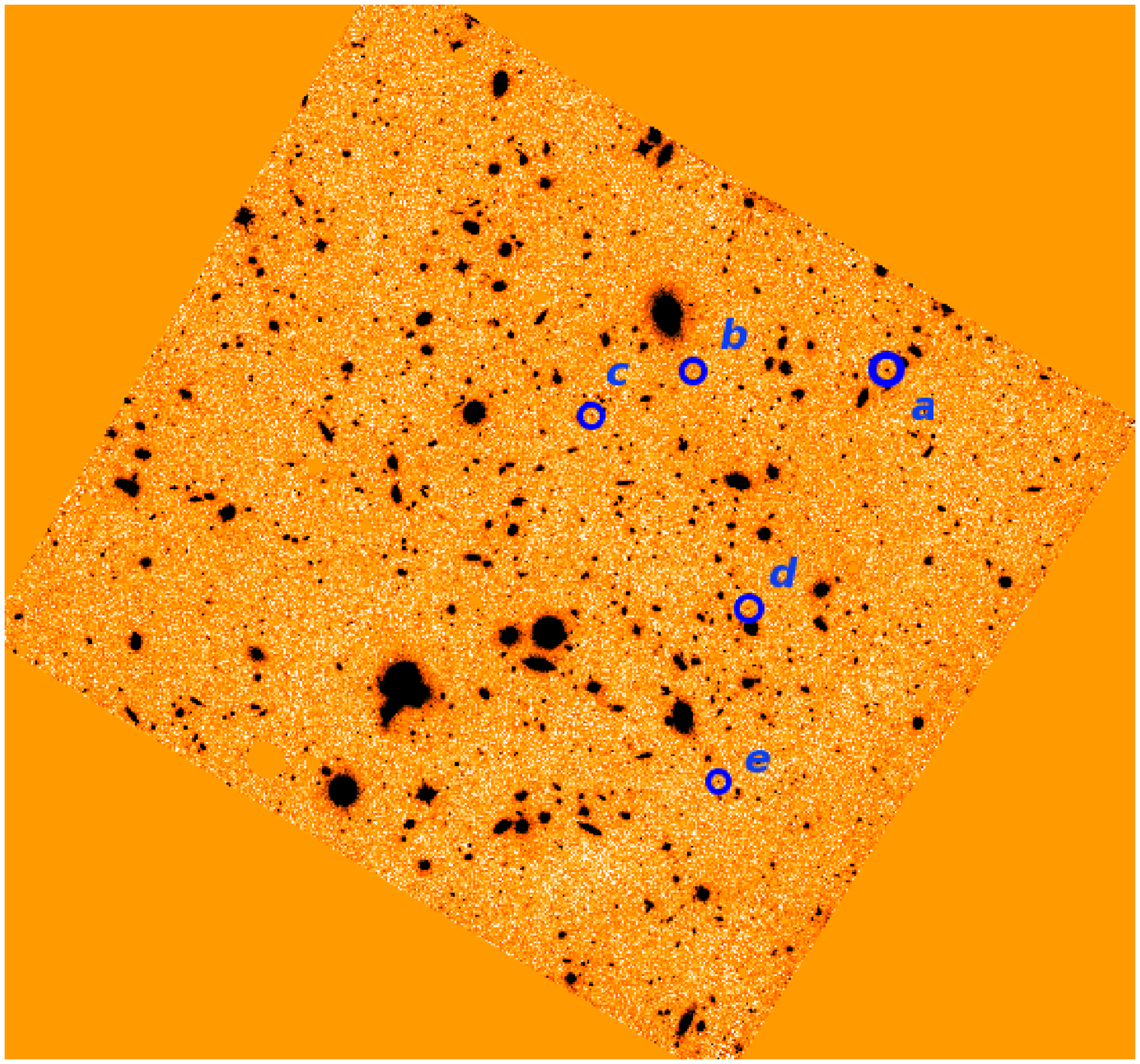}
\includegraphics[scale=0.22]{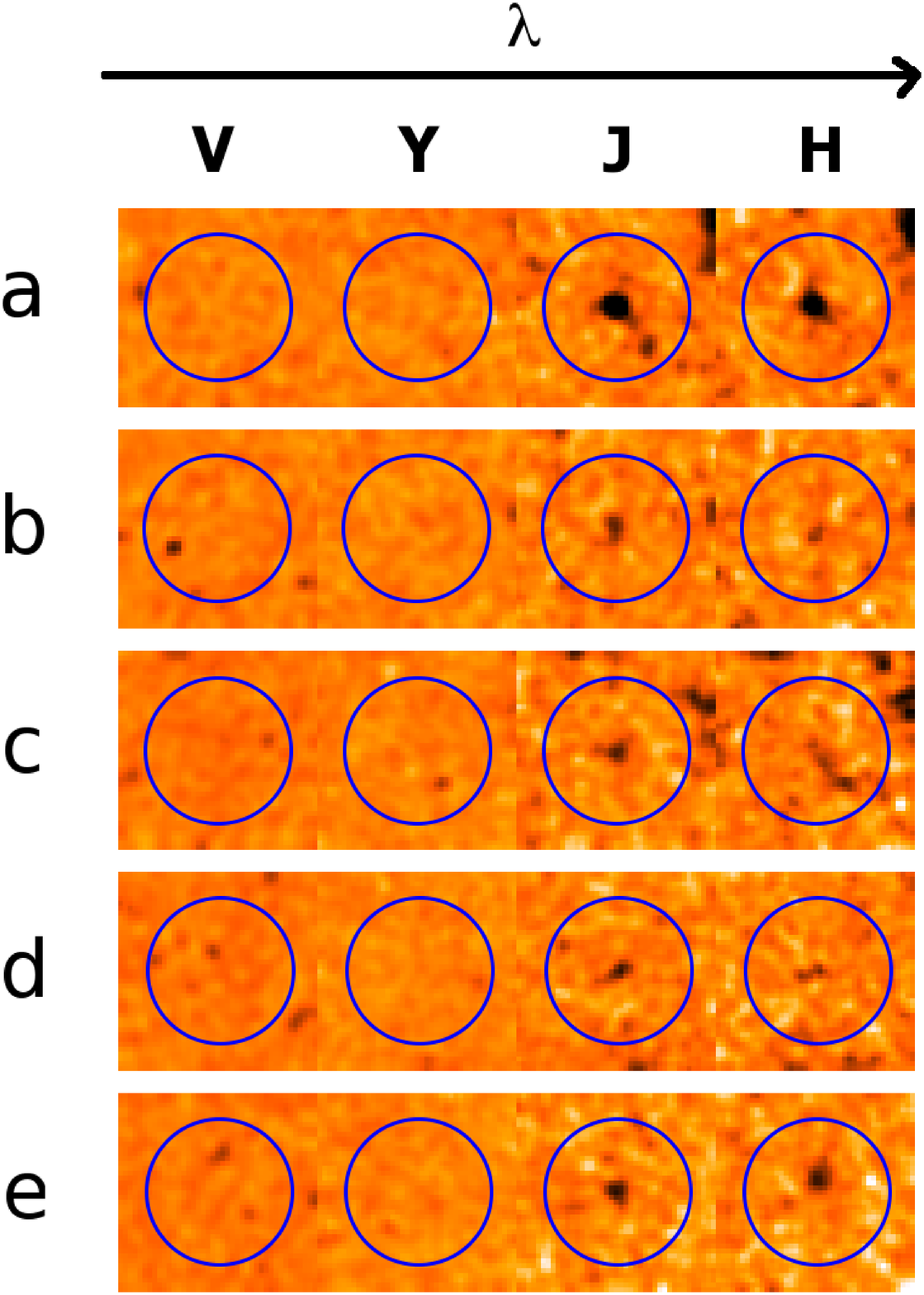}  
\caption{Left panel: $J_{125}$ image of field BoRG58, with
  Y$_{098}$-dropouts indicated by blue circles. Right panel:
  postage-stamp images ($3''.2\times3''.2$) of sources:
  BoRG58\_17871420, BoRG58\_14061418, BoRG58\_12071332,
  BoRG58\_15140953, and BoRG58\_14550613 (top to bottom).}\label{fig:field}
\end{figure}

Five objects within field BoRG58 satisfy the selection criteria for
$Y_{098}$-dropout sources (see Table~\ref{tab:drops}). In addition to
the $m_{J125}=25.8$ source BoRG58\_17871420 discussed in T11, we
identify four additional sources with magnitudes $m_{J125}\sim
27.0$. All candidates are located within a subregion of diameter
$d=70''$ of the $127''\times135''$ WFC3 field
(Figure~\ref{fig:field}). At $z=8$, this separation corresponds to
$3.1$ comoving Mpc. All sources are undetected in $Y_{098}$ and
consistent with the same redshift of $z\sim8$ (Figure~\ref{fig:colcol}
and Table~\ref{tab:drops}). The average $J_{125}-H_{160}$ color of the
four fainter sources (BoRG58\_14061418 - BoRG58\_14550613) appears
marginally bluer ($\Delta(J_{125}-H_{160})\sim-0.2$) than that of the
brightest galaxy (BoRG58\_17871420). About half of this difference is
accounted by PSF effects. The bright dropout is detected over a larger
effective radius ($r_{eff}\sim 0''.33$), hence the difference in
encircled energy between $J_{125}$ and $H_{160}$ is smaller ($0.05$
versus $0.14$ derived for the fainter dropouts).  The remaining
difference after PSF match ($\Delta(J_{125}-H_{160})\sim-0.1)$ could
be due to faint $z\sim8$ galaxies having bluer UV slopes
\citep{bouwens10b}.

As field BoRG58 appears unique among the survey, we performed several
tests with the goal (1) of establishing that the dropouts are real
galaxies rather than noise peaks and (2) of quantifying the
statistical significance of the dropout overdensity in the field.

\subsection{Tests to establish reality of the dropouts}

To demonstrate that the detections of the four fainter sources
BoRG58\_14061418 - BoRG58\_14550613 are real, we performed the
following tests.

{\bf Data quality control:}  We verified that the RMS maps
  around the source locations are regular and do not indicate
  hot/anomalous pixels.

  {\bf Negative image test:} This test allows us to characterize
  non-gaussianity in the tails of the noise distribution and therefore
  to assess the likelihood that detections at $\geqslant 5\sigma$
  confidence are spurious. Using the SExtractor segmentation map, we
  masked sources in our detection image ($J_{125}$), replacing the
  pixels at these locations with pixels sampled randomly from regions
  of the image not containing any source. We then inverted the image
  and proceeded to run SExtractor with the same parameters used for
  the actual science image. In the negative $J_{125}$ image, no
  sources are detected at confidence level $\geqslant
  4.4\sigma$. Hence it is unlikely that the four dropouts
  BoRG58\_14061418 - BoRG58\_14550613 are spurious sources as they are
  detected at $\geqslant 5\sigma$ in the $J_{125}$ science image.

  {\bf H-image random shift test:} Next, we evaluated the likelihood
  that the low-significance detection ($>2.5\sigma$) in $H_{160}$
  arises from noise fluctuations, rather than being intrinsically
  associated to flux from sources detected in $J_{125}$. This is
  especially important in light of the minor misalignment ($0''.15$)
  of BoRG58\_14550613 in these two filters (see
  Figure~\ref{fig:field}). In this test we essentially replace the
  science image in the H band with a noise-only image. Therefore, we
  should expect that the number of sources that have flux in both the
  $J_{125}$ and $H_{160}$ bands at level sufficient to pass the
  $Y_{098}$-dropout selection is much smaller compared to the number
  of sources found when using the H-band science image. For this test,
  we masked sources in the $H_{160}$ image as described above for the
  negative image test. Next, we randomly shifted it along both axes
  using periodic boundary conditions. Then we ran SExtractor and
  constructed source catalogs in all bands, replacing $H_{160}$ with
  the masked and shifted copy. We repeated the experiment 100 times
  with different shifts, determining an expected contamination of
  $N_{contam}=0.03$ objects in the $Y_{098}$-dropouts catalog (a total
  of three objects among the 100 MC realizations, with at most one
  dropout per realization). The probability of detecting four objects
  when $0.03$ are expected is $p<10^{-7}$. This test thus clearly
  shows that the four faint sources surrounding the brighter
  BoRG58\_17871420 have correlated flux in the $J_{125}$ and $H_{160}$
  images, strengthening the likelihood that they are real.

  {\bf Detector-induced noise test:} Next, we evaluated whether
  correlated flux in $J_{125}$ and $H_{160}$ arises from detector
  noise (e.g., charge persistence). The BoRG survey observations have
  been designed so that exposures in $J_{125}$ and $H_{160}$ are
  preceded in the same HST orbit by a $Y_{098}$ exposure of comparable
  length; additional time is devoted to $Y_{098}$ imaging in different
  orbits (T11). Without dithering, a source falls on the same detector
  pixels in every IR filter for exposures in any given orbit. We thus
  combined with \texttt{multidrizzle} the subset of $Y_{098}$
  observations in the two orbits shared with $J_{125}$ and $H_{160}$
  (visits 60 and 62) and measured the $Y_{098}$ flux at the location
  of the dropouts. All dropouts have $S/N<1$ in this $Y_{098}$
  sub-image, in agreement with the non-detection using the full
  combination of all $Y_{098}$ exposures. This demonstrates that the
  detector pixels where the dropout sources are located in $J_{125}$
  and $H_{160}$ are not providing anomalously high electron
  counts. This test also demonstrates that the faint dropouts are not
  spurious sources induced by detector persistence. Otherwise, the
  $Y_{098}$ sub-image would have been affected more than the $J_{125}$
  one (see T11 for a more detailed discussion of our optimization of
  the exposure sequence to avoid persistence-induced spurious
  dropouts). Instrumental origin of the correlated flux in $J_{125}$
  and $H_{160}$ for the faint dropouts is ruled out.

\subsection{Contamination and completeness}\label{sec:complete}

{\bf Improved rejection of contaminants:} We stacked the images
  of the five $Y_{098}$-dropouts identified in the BoRG58 field to
  verify that there is no flux in the $V_{606}$ and $Y_{098}$ bands
  (see Figure~\ref{fig:stack}). Within a circular aperture of radius
  $r=0''.32$, there is indeed no measured flux in these bands, both
  for the stack of the four fainter sources as well as for the stack
  of all five objects. Within the same aperture, the stacked images of
  the four fainter sources have $S/N=8.1$ in $J_{125}$ and $S/N=5.0$
  in $H_{160}$. These values are consistent with the ISOMAG S/N
  measured by SExtractor for the individual sources
  (Table~\ref{tab:drops}). Stacking allows us to set limits that are
  $\sim0.75$ mag deeper than each individual non-detection in
  $V_{606}$ and $Y_{098}$, ruling out to higher confidence the
  possibility that the majority of the dropouts are low-redshift
  contaminants.

  {\bf Dwarf-star contamination:} As discussed in
  Section~\ref{sec:lbg_selection}, our $Y_{098}$-dropout selection is
  highly unlikely to include Galactic ultra-cool dwarfs because we
  require $Y_{098}-J_{125}>1.7$, while these stars have
  $Y_{098}-J_{125}<1.4$ \citep{ryan11}. Furthermore, such
  contamination is even less likely in field BoRG58 compared to a
  typical field in the BoRG survey. In fact, the BoRG58 line of sight
  has a high Galactic latitude ($b_{gal}=59^{\circ}.0$) which is in the
  top $25\%$ of the distribution for $|b_{gal}|$ among the survey
  fields considered here. Finally, we measured size and ellipticity
  for the dropouts and compared them to the PSF. The profiles of both
  the bright dropout itself as well as of a stack of all faint
  dropouts show clear signs of elongation in both J and H bands
  (ellipticity $0.25\pm0.10$), unlike expected for a star (ellipticity
  $\leq0.05$). While the PSF is undersampled in our images due to the
  lack of dithering, the stacks and the bright dropout are also found
  to be wider than the PSF (FWHM$=0.''28\pm 0.''02$ for the sources
  versus FWHM=$0.''218$ for the PSF). Therefore, these tests provide
  further direct evidence that contributes to excluding stellar
  contamination.

\begin{figure} 
\includegraphics[scale=0.8]{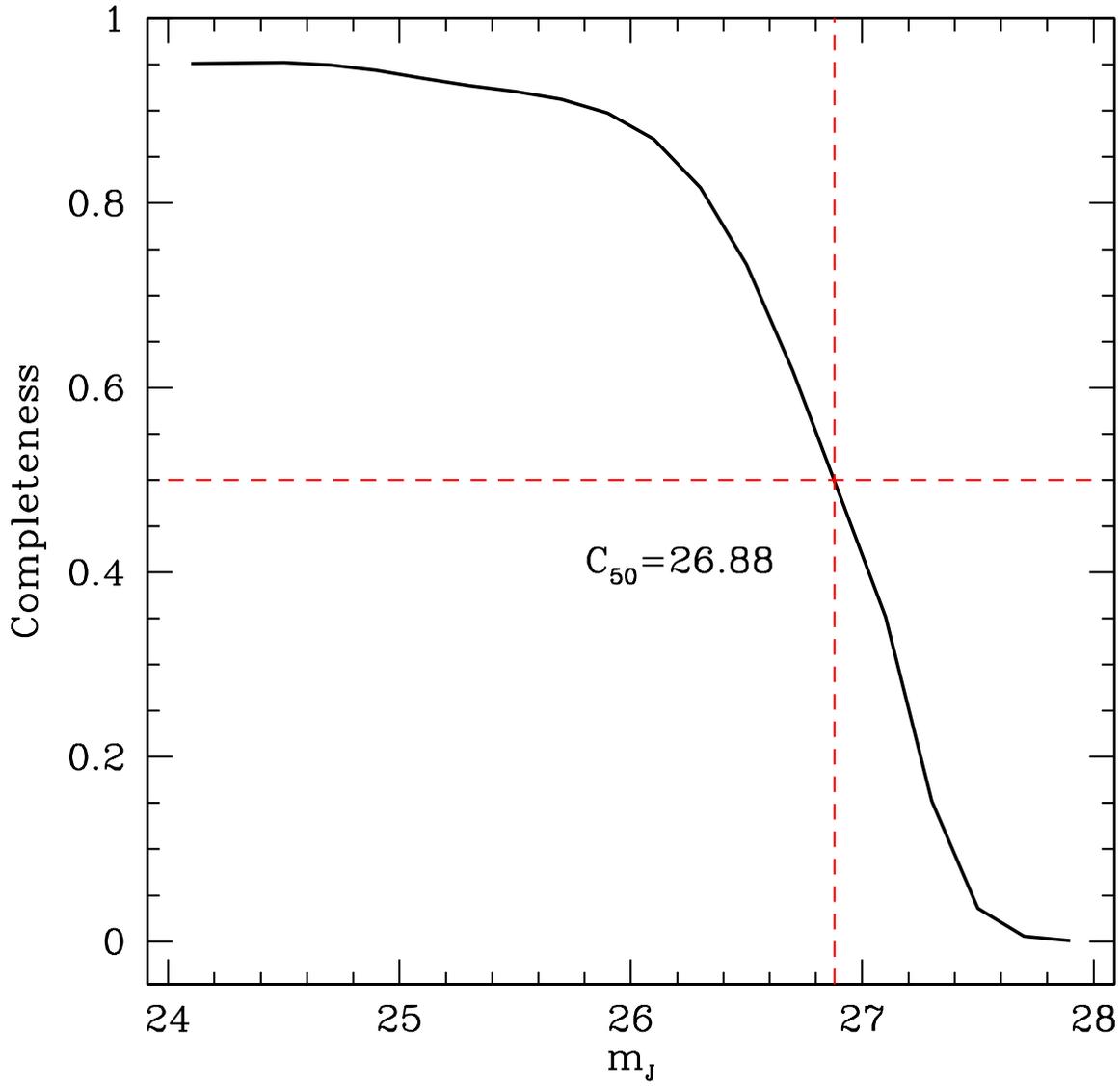}
\caption{Completeness as a function of recovered $m_{J125}$ for the
  $Y_{098}$-dropout selection in field BoRG58. The data are $50\%$
  complete at $m_J=26.88$, estimated using Monte-Carlo experiments of
  source recovery as described in
  \citet{oesch07}.}\label{fig:completeness}
\end{figure}

\begin{figure} 
\begin{center}
\includegraphics[scale=0.44]{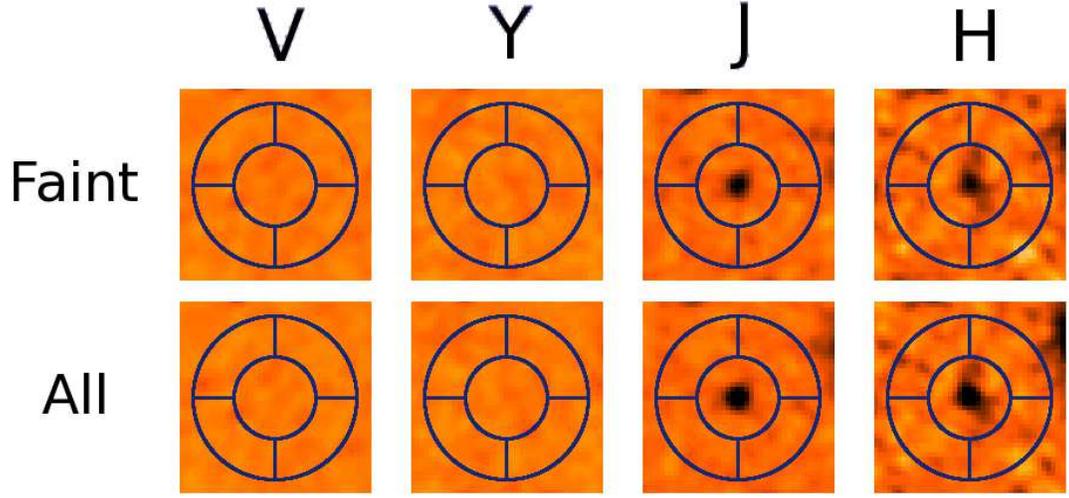}  
\end{center}
\caption{Centered-stacked postage-stamp images ($3''.2\times3''.2$)
 for $Y_{098}$-dropouts in BoRG58 (top: four fainter sources. Bottom:
 all sources). Stacks have no flux in $V_{606}$ and $Y_{098}$, as
 expected for $z\sim8$ sources: $S/N^{(all)}_{V606}=0.9$,
 $S/N^{(all)}_{Y098}=0.6$, $S/N^{(faint)}_{V606}=1.2$,
 $S/N^{(faint)}_{Y098}=1.25$.} \label{fig:stack}
\end{figure}

{\bf Completeness:} We resorted to Monte-Carlo experiments that
  recover artificial sources of varying brightness to evaluate the
  completeness of the data, following the method of
  \citet{oesch07}. Our code evaluates the fraction of input galaxies
  (modeled as Sersic profiles) that are recovered as a function of
  magnitude given our catalog construction and color-color selection
  criteria. We used input galaxies with magnitude $24\leq m_{J125}
  \leq 28$ and a Gaussian UV-slope distribution
  ($\beta=-2.2\pm0.4$). We simulated galaxies in the redshift range
  $6.6\leq z\leq 9.6$ and adopted a log-normal input size distribution
  with mean radius $r=0.''175\times 8/(1+z)$. We find $50\%$
  completeness at $m_{J125}=26.88$ and $>20\%$ completeness at
  $m_{J125}=27.2$, corresponding to the magnitude of our faintest
  candidate (see Fig.~\ref{fig:completeness}).

\subsection{Statistical significance of clustering in BoRG58}\label{sec:sigma}

In this section we quantify the statistical significance of the
clustering of dropouts in field BoR58. First we compare the
overdensity observed against Poisson statistics, which has the
advantage of providing a measure of the clustering significance that
is independent of contamination and completeness as we discuss in
Section~\ref{sec:clustering_survey}. Next, we include the effect of
cosmic variance (galaxy clustering) to assess the overdensity
significance.

As discussed in Section~\ref{sec:clustering_survey}, field BoRG58
contains the highest number of dropouts among the fields
searched. From a simple counts-in-cell statistics considering the full
field of view of WFC3, we derive that an overdensity of $5$ dropouts
or more under the expectation of $\langle N\rangle=21/23=0.913$ is
realized with probability $P(N\geq5)=0.0025$ under Poisson
statistics. As field BoRG58 was identified in T11 as the one
containing the best and most robust candidate in the survey, therefore
establishing an a priori expectation for clustering, we are justified
in considering the dropout overdensity significant at $>99.75\%$
confidence. If we were to consider a more conservative measure and
consider the probability of finding one overdensity with $N\geq5$
$Y_{098}$-dropouts among the four fields with a bright candidate from
T11, we would still obtain that the dropouts are clustered at
$>99\%$ confidence. 

Furthermore, the dropouts in BoRG58 are located within a subregion of
the WFC3 field of view; therefore the overdensity is even more
significant. To correctly model the finite field of view and edge
effects, we resorted to a Monte-Carlo simulation to quantify the
probability that five sources are located within a region with
$d=70''$ within the $127''\times135''$ WFC3 field if they are
distributed uniformly with $\langle N \rangle = 0.913$ per full WFC3
field. Such configuration happens with probability
$p=7.2\times10^{-5}$ for field BoRG58, or with probability
$p=2.88\times10^{-4}$ under a more robust measure of requiring one
overdensity among the four fields with bright dropouts. In both cases,
the significance of the overdensity is increased by a factor $\sim 35$
compared to the counts-per-field estimate.

The comparison with Poisson statistics demonstrates, to very high
confidence level ($p>0.999932$), that the overdensity of sources in
field BoRG58 is physical rather than originating from random
fluctuations. The distribution of galaxy counts in pencil-beam surveys
differs from Poisson statistics because of clustering (cosmic
variance), as measured from the two-point correlation function
\citep{peebles80}. Therefore we take into account cosmic variance to
evaluate whether the overdensity of galaxies in BoRG58 is likely
associated to a protocluster of galaxies, with the (majority of)
dropouts in a narrow redshift range. The alternative scenario is a
fluctuation in the number counts that is made more probable because of
projected galaxy clustering from sources that are at different
redshift within the $Y_{098}$-dropout selection window.

As discussed in Section~\ref{sec:clustering_survey}, the
$Y_{098}$-dropout sample has the highest bias (that is strongest
clustering) in the case where the survey is complete and there is no
contamination. We recall that this is because an incomplete survey
samples lower halo masses and because contaminants and $z\sim 8$
dropouts have no cross-correlation. If we assume a mean number density
of $\langle N \rangle =0.913$ $z\sim 8$ candidates per WFC3 field, we
derive from our cosmic variance calculator a bias\footnote{Note that
  the bias of the population of contaminants is significantly lower
  than the bias of $z\sim8$ sources with the same luminosity. From the
  conditional luminosity function model of \citet{cooray05}, a
  contaminant at $z\sim2$ with observed magnitude $m_{J125}\sim26.5$
  ($M_{J125}\sim-18.3$) is expected to live in a dark-matter halo of
  mass $M_{halo}\approx 10^{11}~\mathrm{M_{\sun}}$ with bias $b\approx
  1.5$.}  $b=8.9$, that is a cosmic variance of $48\%$
\citep{ts08}. This estimate relies on an abundance match of
high-redshift galaxies with dark-matter halos, which is the working
assumption adopted by several studies
\citep{newman02,somerville04,ts08,munoz10,robertson10}. This modeling
has been tested observationally at $z\lesssim 6$ by determination of
the two-point correlation function of Lyman-break galaxies
\citep{overzier06,lee09}, and indirectly at $7\lesssim z \lesssim 10$
because conditional luminosity function models correctly predict the
observed abundance of dropout galaxies in the HUDF and ERS/CANDELS
surveys \citep{trenti10,oesch11}. In any case, the uncertainty in the
bias of the Y-dropout sample propagates only mildly into the final
determination of the number counts fluctuations. This is because
cosmic variance and Poisson dispersion are summed in quadrature;
therefore, when the expected number of objects per field is smaller or
comparable to unity, Poisson uncertainty dominates the total
dispersion. In fact, if we assume a pencil beam of $d=70''$ with
redshift selection $7.5\leq z\leq 8.5$ (T11), we derive a variance of
$0.53$ $Y_{098}$-dropouts per cell of diameter $d=70''$ around their
average value of $0.20$. That means that the overdensity of five
$Y_{098}$-dropouts in BoRG58 is significant at $8.9\sigma$, after
including the effects of galaxy clustering. For reference, under
Poisson statistics the overdensity would have been significant at the
$10.1\sigma$ level.

\begin{figure}
  \plotone{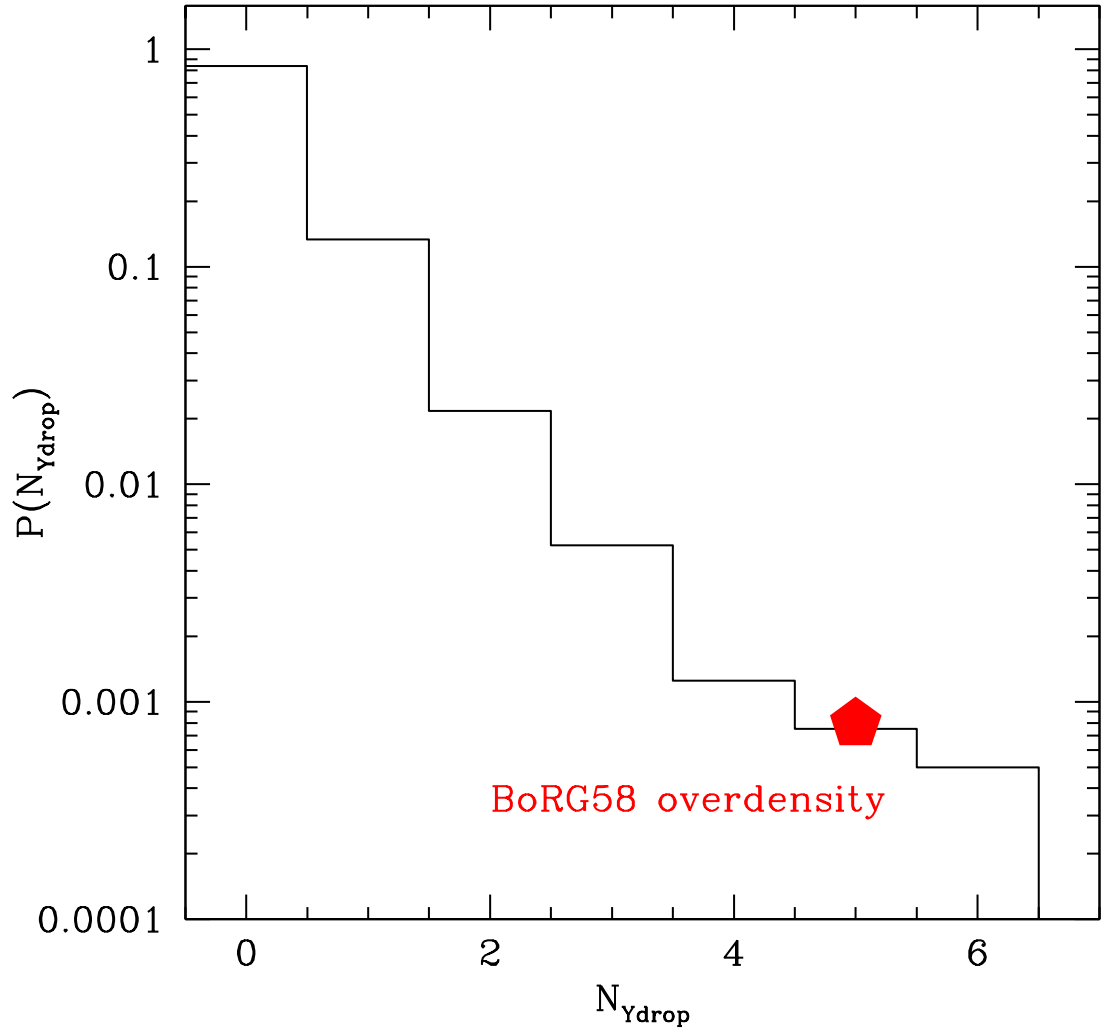}\caption{Number count distribution
    of $Y_{098}$-dropouts in a pencil beam of comoving volume
    $2.83\times 2.83 \times 311 ~\mathrm{Mpc^3}$ (corresponding to
    $62''\times62''\times$$\Delta z$, with $ \Delta z= 1.0$ at
    $\langle z \rangle =8.0$), traced through a cosmological
    simulation with box edge $l_{box} = 143~\mathrm{Mpc}$ (comoving)
    and $N_p=512^3$ DM particles \citep{ts08}. Only $5$ out of $4000$
    random realizations of the pencil beam contain an overdensity of
    at least $5$ sources. In four of the five cases the overdensity is
    associated with a 3D proto-cluster. A random superposition of
    $Y_{098}$-dropouts at different redshifts along the pencil beam
    happens with probability $p=1/4000=2.5\times 10^{-4}$, thereby
    suggesting a physical origin for the BoRG58 overdensity to a high
    degree of confidence.}\label{fig:clustering_cosmosim}
\end{figure}

An approximate analytical form for the number counts distribution in
presence of cosmic variance could be derived analytically under the
same assumption that light traces dark matter (e.g.,
\citealt{adelberger98,robertson10}), but that would not allow us to
distinguish excess number counts originating from physical
overdensities versus line-of-sight random superpositions. Therefore,
we traced a pencil beam trough a cosmological simulation, following
the method described in \citet{ts08} to construct the full probability
distribution of the expected number counts (see also
\citealt{munoz10}). We used the box of edge $l_{box} =
100~\mathrm{Mpc}~{h^{-1}}$ ($h=0.7$) simulated with $N_p=512^3$ DM
particles as discussed in \citet{ts08}.
Snapshots and halo catalogs were constructed every $\Delta z =0.125$,
allowing us to model the redshift evolution of the box over multiple
snapshots as the $Y_{098}$-dropout pencil beam is traced ($\Delta z =
1$ centered at $z=8$). We set a minimum threshold of 92 particles for
the dark-matter halos to be included in the pencil beam, as this gives
an average number of counts in a WFC3-like pencil-beam that is equal
to the observed value of $\langle N_{Ydropout} \rangle =0.913$ (every
dark-matter halo above the threshold is populated with a galaxy). We
then considered a pencil beam equivalent in volume to the $d=70''$
region that contains the BoRG58 overdensity, but shaped as a
parallelepiped for computational reasons: $2.83\times 2.83 \times 311
~\mathrm{Mpc^3}$ (corresponding to $62''\times62''\times$$\Delta z$,
with $ \Delta z= 1.0$ at $\langle z \rangle =8.0$). We ran 4000 random
realizations of the pencil beam, obtaining $\langle N \rangle =0.2$
and variance $0.53$, in agreement with the analytical estimate
discussed above. The full number count distribution of the
$Y_{098}$-dropout counts is presented in
Figure~\ref{fig:clustering_cosmosim}. Five realizations in $4000$ have
$N\geq 5$. Only one of these five fields with high number counts
arises because of superposition of dropouts at different redshift
along the pencil beam. The remaining realizations have all dropouts
within a region of comoving size $5~\mathrm{Mpc}$.  This numerical
experiment allows us to set a probability $p=1-1/4000=0.99975$ that
the BoRG58 overdensity is a not a result of galaxy clustering along
the line of sight. In addition, in the case of random superposition
along the line of sight, all dark-matter halos have comparable masses
near the low-mass cutoff to be included in the beam. In contrast, the
fields that include 3D clusters have at least one halo with mass about
four times higher than the minimum mass. Under the assumption that
luminosity correlates with mass, the presence of a very luminous
galaxy as in the case of field BoRG58 further strengthens the
likelihood of a proto-cluster scenario.

Finally, we note that the overdensity of $Y_{098}$-dropout sources
observed in BoRG58 is also more significant than any overdensity of
$z\sim8$ sources found among the 10 WFC3 pointing in the GOODS/ERS
fields by \citet{bouwens10c} who found two $z\sim8$ candidates in a
region of diameter $d=74''$.

To summarize, all tests performed imply that the $Y_{098}$-dropout
overdensity in BoRG58 is real and has high statistical significance.
To investigate in more detail a possible scenario for the physical
origin of the overdensity, we resort to numerical modeling of galaxy
assembly during the epoch of reionization in the next Section.

\section{Theoretical/numerical modeling of the
  overdensity}\label{sec:theory}

Under the assumption that the five $Y_{098}$-dropout of field BoRG58
are at $z\sim8$, we can infer the dark-matter properties of the halos
in which they reside by resorting to the Improved Conditional
Luminosity Function model of \citet{trenti10}. From their halo-mass
versus galaxy-luminosity, we infer that the brightest member of the
overdensity ($M_{AB}=-21.3$) lives in a $M_h\approx(7\pm2)\times
10^{11}~\mathrm{M_{\sun}}$ halo. Such halo is a $5\sigma$ density peak
with comoving space density $\approx9\times10^{-7}~\mathrm{Mpc^{-3}}$,
which has approximately a $20\%$ chance of being present within the
volume probed by the BoRG survey ($2.3\times
10^{5}~\mathrm{Mpc^3}$). This mild tension could be eased if there is
some redshift evolution of the halo-mass versus galaxy-luminosity
relation so that galaxies at higher $z$ become brighter at fixed halo
mass as suggested by \citet{munoz10b}; the \citealt{trenti10} model is
calibrated at $z=6$ while the Y-dropouts are at $z\sim 8$. If this is
the case, we can derive instead the halo mass from number-density
matching, obtaining a slightly lower value
($M_h\sim4\times10^{11}~\mathrm{M_{\sun}}$). The fainter galaxies
($M_{AB}\approx-20.0$) are expected to be hosted in lower mass, and
more common, halos with $M_h\approx10^{11}~\mathrm{M_{\sun}}$
($4\sigma$ peaks).

Using extended Press-Schechter modeling \citep{munoz08}, we expect on
average $N=4.8$ halos with $M_h\geq10^{11}~\mathrm{M_{\sun}}$
surrounding BoRG58\_17871420 within a sphere of diameter
$d=3.1~\mathrm{Mpc}$ (comoving) corresponding to $d\leq70''$.  For
reference, the average number of halos above this mass threshold for a
random region of the universe of the same volume at $z=8$ is $\langle
N\rangle<10^{-3}$. A massive dark matter halo like the one expected
for BoRG58\_17871420 boosts by more than three orders of magnitude the
number density of halos that can host galaxies sufficiently luminous
to be detected. Similar results are given by a more powerful
clustering model originally developed to describe the Shapley
Supercluster \citep{munoz08b}. This analytical estimate shows that
clustering of fainter dropouts around a very bright dropout is
naturally expected. The prediction can be extended further to argue
that if no clustering is observed at all, then it is likely that the
very bright dropout is not at $z\sim 8$, but rather a foreground
contaminant, because in this case the source is within a dark-matter
halo with lower bias (see Section~\ref{sec:sigma} and also
\citealt{munoz08}). We note that, as discussed in
Section~\ref{sec:clustering_survey}, all the bright dropouts in the
BoRG survey show evidence of clustering. The signal is not as strong
as in the case of BoRG58, but is still significant at $>99.84\%$
confidence.

\begin{figure}
 \plottwo{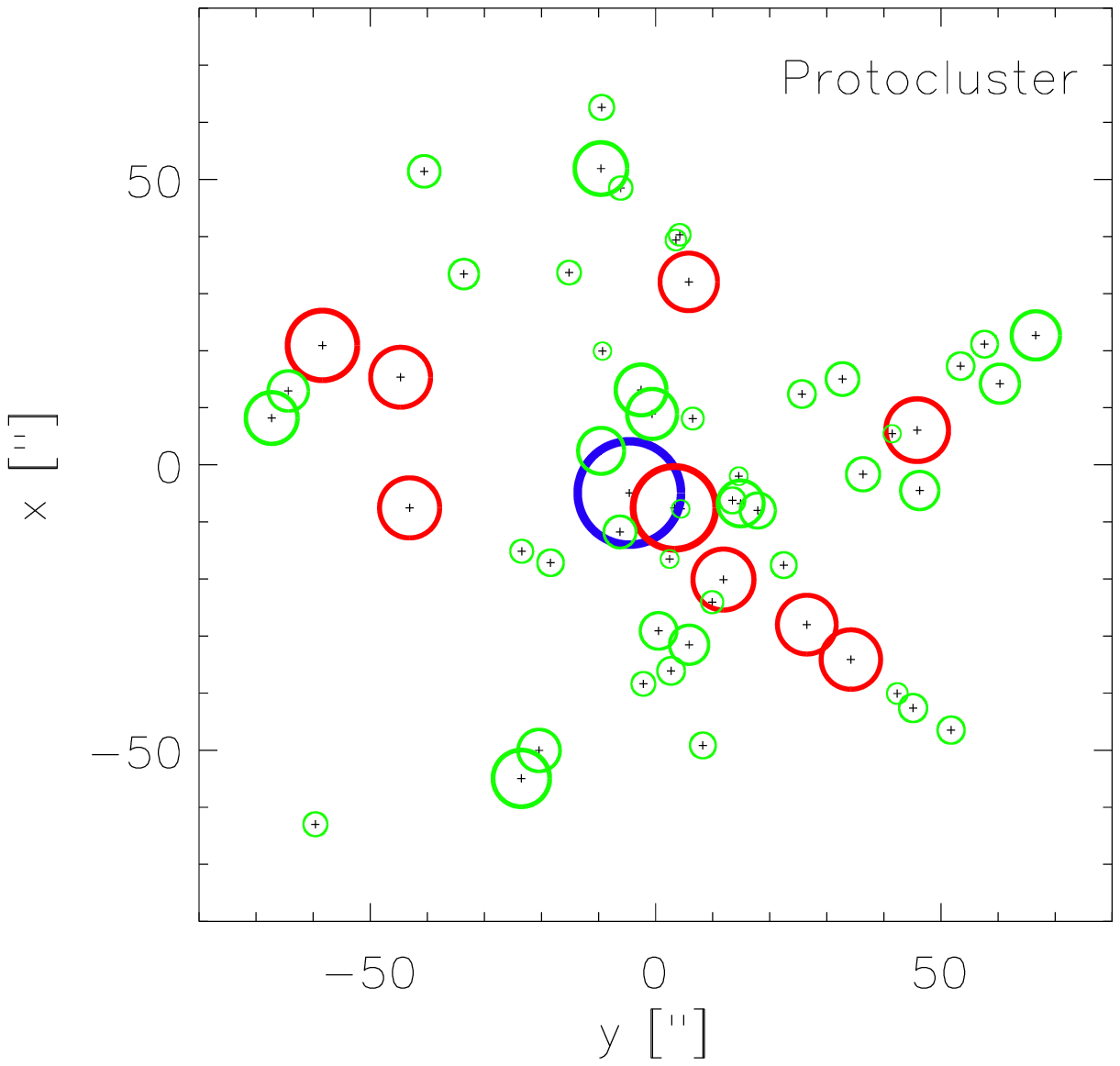}{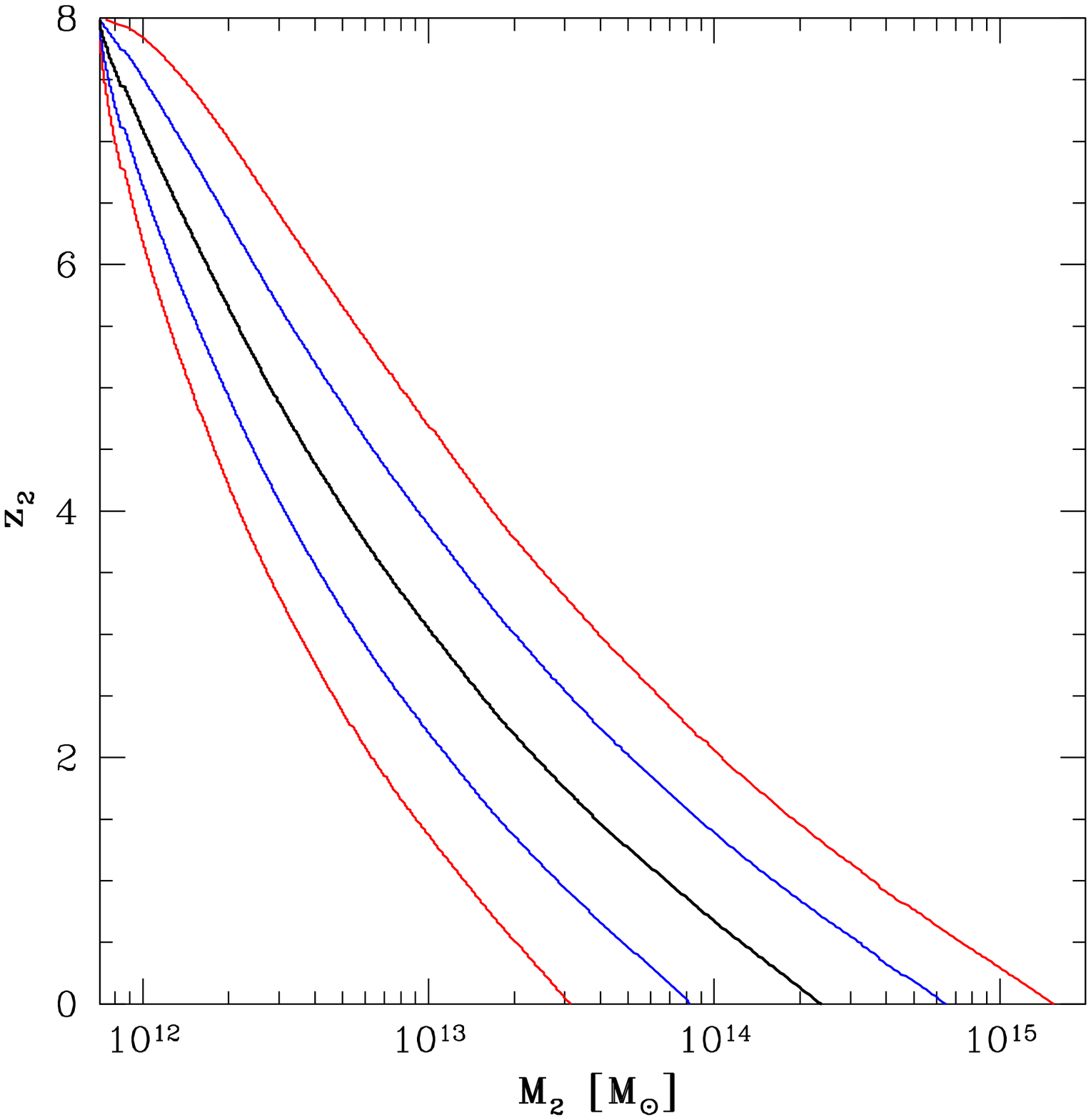} \caption{Left:
   Dark-matter halo distribution at $z=8.08$ for simulated
   proto-cluster (comoving volume
   $11\times11\times19~\mathrm{Mpc^3}$; from
   \citealt{romanodiaz10}). $M_h\gtrsim
   10^{11}~\mathrm{M_{\sun}}$ halos (red and blue circles) host
   galaxies likely detectable in the BoRG58 field at the current depth (the blue circle denotes
   the most massive halo in the simulation). Green circles are less massive halos
   ($10^{10}~\mathrm{M_{\sun}} \leq M_h\leq
   10^{11}~\mathrm{M_{\sun}}$). 
Right: BoRG58 overdensity evolution predicted from extended
Press-Schechter modeling. Black line: median of probability that a
$z=8$ halo with $M_h=7\times 10^{11}~\mathrm{M_{\sun}}$ evolves into
a halo of mass $M_2$ at $z_2$ (from \citealt{tss08}). Blue/red lines
are $1\sigma$ and $2\sigma$ confidence contours. 
}\label{fig:sim}
\end{figure}

Analytical predictions of clustering are limited to spherical volumes,
whereas observations select galaxies within a pencil beam
\citep{ts08,munoz10}. For more accuracy, we use a set of $10$
cosmological simulations designed to study high-$z$ galaxy formation
in overdense environments \citep{romanodiaz10}. These runs follow only
the evolution of dark matter and similar to the cosmic variance
estimates constructed in this paper rely on the assumption that light
traces matter\footnote{In passing, we note that the conclusion of
  \citet{romanodiaz10} was that the predicted clustering of
  Lyman-break galaxies around the brightest $z\sim 6$ QSOs was
  stronger than observed, under the assumption that all these QSOs
  live in dark matter halos with masses inferred from abundance
  matching ($M_h>10^{12}~\mathrm{M_{\sun}}$). Therefore it was
  proposed there that the tension with the observations could be eased
  if some of the QSOs are instead hosted in less massive
  halos. Recently, \citet{volonteri11} showed that, because of the SDSS
  magnitude-limited selection, this is likely to be the case.}. The
simulations have a comoving volume $(28.6~\mathrm{Mpc})^3$ and mass
resolution $m_p=3\times10^8~\mathrm{M_{\sun}}$. Halos with
$10^{11}~\mathrm{M_{\sun}}$ are thus well resolved with several
hundred particles \citep{trenti10sim}. The initial conditions were
constrained to contain a $M_h \sim 10^{12}~\mathrm{M_{\sun}}$ halo at
$z=6$, imposing the input overdensity at the center of the box with
the \citet{hoffman91} method. These simulations have a central
dark-matter halo with $\langle M_h(z=8.08)\rangle
\sim5\times10^{11}~\mathrm{M_{\sun}}$ at $z=8$ and are therefore
appropriate for investigating the environment of
BoRG58\_17871420. Within a $70''\times70''$ field of view with a
line-of-sight extension of $19~\mathrm{Mpc}$ ($\Delta z\approx0.05$),
there are on average $N\sim 6.4$ halos with $M_h\geqslant
10^{11}~\mathrm{M_{\sun}}$.  We find $N=10$ in the realization with
the highest abundance of halos (Figure~\ref{fig:sim}). For comparison,
we expect $N=0.013$ halos in such volume for a random region of the
universe at $z=8$. Numerical simulations by \citet{overzier09} show
that projected overdensities of dropouts might not be associated with
actual 3-D overdensities, but the probability is negligible for the
field of view considered here, as we demonstrated from our analysis of
numerical simulations in Section~\ref{sec:sigma} ($p=2.5\times
10^{-4}$).

These theoretical expectations indicate that, if the bright $z\sim 8$
candidate in field BoRG58 lives in a massive dark-matter halo, then
the overdensity of fainter $Y_{098}$-dropouts in its proximity is
naturally explained as a proto-cluster structure. Although without
spectroscopic confirmation we cannot demonstrate that the overdensity
is composed of $z\sim 8$ galaxies at a common redshift, all other
hypotheses (either contamination from foreground galaxies or Galactic
stars, or $7.5\lesssim z\lesssim8.5$ galaxies at different redshifts)
are highly unlikely (Section~\ref{sec:sigma}). Of course, the current
data cannot exclude that some of the faint dropouts are at different
redshifts with respect to the bright dropout (including being
foreground contaminants). Our numerical simulations indicate that
deeper data would not only confirm the nature of the currently
identified $Y_{098}$-dropouts, but more importantly they would
discover additional, fainter candidates living in the proximity of the
massive dark matter halo that hosts BoRG58\_17871420.

Assuming that the BoRG58 overdensity is composed of galaxies at the
same redshift, our modeling gives us insight on the fate of the
overdensity as it evolves with time. If the five galaxies are within a
sphere of radius $r=1.5$ comoving Mpc, then the region is already
non-linear, with an average overdensity
$\Delta\rho/\langle\rho\rangle>10$ by counting just the mass within
the galaxy host halos. From our cosmological simulations, we derive
that these galaxies are likely gravitationally bound and will collapse
by $z\sim3$ into a single dark-matter halo (see Figure~7 in
\citealt{romanodiaz10}). This is confirmed by extended Press-Schechter
modeling, which takes into account the effects of halo growth and
clustering, allowing us to derive the probability
$p(M_h>M_2,z_2|M_1,z_1)$ that a dark-matter halo of mass $M_h=M_1$ at
redshift $z_1$ will evolve into a halo with mass $M_h >M_2$ at
redshift $z_2<z_1$ \citep{lacey93}. We used the extended
Press-Schechter code developed in \citet{tss08} to derive the expected
mass distribution for the descendants of a BoRG58-like halo (mass
$M_h(z=8)=7\times10^7~\mathrm{M_{\sun}}$) at lower redshift, as shown
in right panel of Figure~\ref{fig:sim}. The figure predicts that the
BoRG58 overdensity will evolve into one of the first massive galaxy
clusters formed in the universe, with $M_h>2\times
10^{14}~\mathrm{M_{\odot}}$ by $z=0$.

\section{Summary and Conclusions}\label{sec:conclusion}

Assuming a correlation between galaxy luminosity and dark-matter halo
mass \citep{vale04,cooray05,trenti10}, a natural prediction of
dark-matter clustering is that the brightest galaxies at $z>6$ should
be surrounded by an overdensity of fainter galaxies at similar
redshift \citep{munoz08}. In this paper we tested this scenario using
multi-band HST data from the BoRG survey (T11). Initial results from
the survey identified four bright $z\sim 8$ candidates as
$Y_{098}$-dropout sources from a large area medium-deep survey in four
bands (V, Y, J, H), covering $130$ arcmin$^2$ to a median $5\sigma$
sensitivity of $m_{AB}\sim 27$. As the four bright candidates in T11
are detected at high confidence ($S/N>8$), in this paper we searched
for fainter dropouts (detected at $5<S/N\leq 8$) with the goal of
testing whether the faint sources are found preferentially in the
fields containing bright candidates. Our extended search (see
Section~\ref{sec:clustering_survey} and Table~\ref{tab:field_summary})
identified 17 additional dropouts within 23 independent fields. Eight
of the new sources are located within the four fields that contain a
bright $Y_{098}$-dropout identified by T11. Each of these fields
contains at least one faint dropout. By means of Monte-Carlo
simulations we established that there is a random chance $p=0.0016$
that this occurs if no correlation between bright and faint dropouts
is present. Therefore, the theoretical expectation that such
correlation is present is verified from the BoRG survey data at
$99.84\%$ confidence. This result holds independent of the
contamination rate and completeness of the $Y_{098}$-dropout
selection. Furthermore, contamination and incompleteness reduce the
strength of such correlation
(Section~\ref{sec:clustering_survey}). Therefore, our result is robust
against these effects. The presence of the correlation suggests that
the samples of both bright and faint dropouts are not severely
affected by contamination, as this would weaken the strength of the
cross-correlation signal.

Interestingly, field BoRG58, in which T11 identified the best $z\sim
8$ candidate of the BoRG survey based on the photometric data, shows
the strongest overdensity of faint $Y_{098}$-dropouts: four additional
sources are identified in this paper. The other three fields
containing a bright candidate from T11 still show an excess of fainter
sources compared to the typical BoRG survey number counts, but their
overdensities are more modest (one or two faint dropouts per
field). There can be several contributing factors that lead to the
difference between BoRG58 and the other fields with a bright
candidate. First, the bright dropout in field BoRG58 is the one with
the highest signal-to-noise detection ($S/N_{J125}=13.2$). Therefore,
this gives the broadest dynamic range among the fields with bright
candidates to search for fainter companions with
$S/N_{J125}>5$. Another possibility is that this is due to scatter in
the halo-mass versus galaxy luminosity relation, typically modeled as
a log-normal distribution \citep{cooray05}. In this case, the bright
dropout in field BoRG58 might live in a halo with (above) average
mass, while the other bright candidates might be over-luminous for
their halo mass (and hence not clustered as strongly). This scenario
could be favored because of the observational bias toward outliers
present in a magnitude-limited sample, in analogy to what has been
discussed in the context of the $z\sim 6$ QSOs luminosities by
\citet{volonteri11}. Finally, the other very bright $Y_{098}$-dropout
of the sample, a $m_{AB}=25.5$ galaxy in field BoRG1k, may be a
contaminant, as discussed in Section~5.2 of T11. In this respect, it
is interesting to note that the next best $z\sim 8$ candidate
identified by T11 is in field BoRG0t, which contains the second most
significant overdensity of faint $Y_{098M}-$dropouts, with two
additional sources detected in addition to the $m_{AB}=26.7$ $Y_{098}$
galaxy discussed in T11. This galaxy is almost one magnitude fainter
than the one in BoRG58, so it is not surprising to find fewer members
in the overdensity.

Given the special interest of field BoRG58, we carried out extensive
analysis and data interpretation. In Section~\ref{sec:tests}, we
quantified by means of extensive testing that the overdensity of five
dropouts in field BoRG58 is significant at $99.97\%-99.999\%$
confidence under Poisson statistics, with the range depending on
{\it{a priori}} assumptions. As in the case of the global analysis for
correlation among counts in the survey, this result is independent of
contamination and completeness. We also quantified the impact of
galaxy clustering in the determination of the statistical significance
of the overdensity. Under the most conservative scenario of a complete
survey with no contamination (this choice maximizes the bias of the
$Y_{098}$-dropout population), we derive that the overdensity of
dropouts in BoRG58 lies $8.9\sigma$ away from the average number
count. To quantify the exact probability associated with this we
constructed the number count distribution by tracing a pencil beam
through a cosmological simulation as described in \citet{ts08}. We
derive that five counts as in the BoRG58 overdensity are associated with
a clustered structure at a common redshift with probability $p\geq
99.975\%$.

These results strengthen our confidence that the bright
$Y_{098}$-dropout BoRG58\_17871420 is indeed at $z\sim8$ and lives in
a dark-matter halo of mass $M_h\sim (4-7)\times
10^{11}~\mathrm{M_{\sun}}$ ($5\sigma$ density peak), rather than being
a lower redshift contaminant. If that galaxy were at $z\sim1.5-2$, it
would be a low-luminosity galaxy hosted in a common $\lesssim 2\sigma$
peak which would not be surrounded by an overdensity of similar
sources \citep{munoz08}.  Further evidence for the $z\sim 8$ nature of
the source is given by Keck-NIRSPEC spectroscopy
\citep{schenker11}. While no Ly$\alpha$ emission line has been
detected, to a median equivalent width of $EW \gtrsim
40~\mathrm{\AA}$, this spectroscopic follow-up rules out that the
galaxy is a foreground contaminant at $z\sim 2$ with flux in the J and
H bands primarily contributed by strong emission lines, which would
have been detected in the spectrum at very high confidence
($>20\sigma$).

The faint $Y_{098}$-dropouts in BoRG58 are however at the detection
limit of the current data and likely affected by significant
contamination and incompleteness (Section~\ref{sec:complete}). To
confirm beyond reasonable doubt the existence of a physical 3D
overdensity, and to explore in more detail its properties, new
observations are needed. Deeper HST data, both in the optical and in
the near-IR would improve the rejection of contaminants and likely
discover new members of the overdensity, as expected from our
theoretical/numerical model of the overdensity
(Section~\ref{sec:theory}). In this respect, we note that our catalog
contains several sources in the field that are candidate
$Y_{098}$-dropouts but are only detected in the J band at $4 <
S/N_{J125} < 5$ confidence. In addition, Spitzer/IRAC observations
($3.6~\mathrm{\mu m}$ and $4.5~\mathrm{\mu m}$) would allow us to
exclude to higher confidence dusty and passively evolving $z\sim1.5$
contaminants, and more importantly to derive the age and stellar mass
for the brightest dropout, BoRG58\_17871420.  These observations could
rule out that this ultra-bright $z\sim8$ candidate is a young
($t\lesssim5~\mathrm{Myr}$) and not very massive starburst, a scenario
which we did not consider in our theoretical modeling of the
overdensity. In fact, typical $z\sim8$ sources have ages of
$t\sim100-300~\mathrm{Myr}$ (see~\citealt{labbe10}).

Finally, it will be possible to further investigate the nature of the
BoRG58 overdensity with the next generation of ground-based,
multi-object spectrographs. The Keck observations of
\citet{schenker11} are only moderately deep, with five hours of
integration time split in two different filters to cover the expected
wavelength range of Ly$\alpha$ emission (as a reference,
\citealt{lehnert10} observed with the VLT SINFONI spectrograph a
$z\sim 8$ galaxy for about 15 hours in a single filter). Galaxies at
$z\sim 8$ are likely to have smaller Ly$\alpha$ equivalent width
compared to lower redshift objects. Theoretical predictions suggest
that Ly$\alpha$ emission should be present with $EW\gtrsim 15
\mathrm{\AA}$ \citep{dayal11} and that some Ly$\alpha$ flux should be
transmitted even in case the IGM is neutral because of radiative
transfer induced by galactic outflows \citep{dijkstra10}. This limit
is within reach of a deeper spectroscopic follow-up from the ground
and especially with a multi-object spectrograph such MOSFIRE at Keck,
which could provide the definitive evidence that we are witnessing the
assembly of one of the first galaxy clusters in the Universe.

\acknowledgements 

We kindly thank an anonymous referee for helpful suggestions that
improved the paper, Matthew Schenker for sharing the sensitivity curve
of his spectroscopic follow-up of the bright $z\sim 8$ candidate in
field BoRG58, and Brant Robertson for useful discussions. This work
was supported in part through grants HST-GO-11700 and HST-AR-12639
provided by NASA through a grant from STScI, which is operated by
AURA, Inc., under NASA contract NAS 5-26555.  We acknowledge support
from the University of Colorado Astrophysical Theory Program through
grants from NASA (NNX07AG77G) and NSF (AST07-07474,
AST08-07760). Support for this work was provided by NASA through
Hubble Fellowship grant HF-51278.01 awarded to PO by the Space Telescope
Science Institute, which is operated by the Association of
Universities for Research in Astronomy, Inc., for NASA, under contract
NAS 5-26555. ERD acknowledges support from the SFB 956 "Conditions and
Impact of Star Formation" by the Deutsche Forschungsgemeinschaft
(DFG).




\clearpage

\newpage
\begin{deluxetable}{lrrrr}
  \small \tablecolumns{5} \tablewidth{0pt} \tablecaption{$Y_{098}$-dropout ($z\sim8$
    candidates) in the BoRG survey.\label{tab:field_summary}}
  \tablehead{\colhead{Field}& \colhead{$m_{J125}$} & \colhead{$S/N_{J125}$} &
    \colhead{R.A.} &\colhead{Decl.}}  \small \startdata
  BoRG70 & 26.4 & 5.2 &  157.7291 & +38.0474 \\
  BoRG66 & 26.2 & 8.7 &  137.2732 & -0.0297 \\
  BoRG66 & 26.5 & 7.0 &  137.2879 & -0.0338 \\
  BoRG58 & 25.8 & 13.0 & 219.2107 & +50.7260 \\
  BoRG58 & 27.2 & 5.1 & 219.2241 & +50.7260 \\
  BoRG58 & 26.9 & 5.5 & 219.2311 & +50.7241 \\
  BoRG58 & 27.2 & 5.4 & 219.2203 & +50.7156 \\
  BoRG58 & 27.0 & 6.0 & 219.2224 & +50.7081 \\
  BoRG2t & 26.6 & 6.8 & 95.9036 & -64.5480  \\
  BoRG1v & 26.4 & 5.2 & 187.4776 & +7.8286  \\
  BoRG1k & 25.5 & 11.4 & 247.8968 & +37.6039 \\
  BoRG1k & 26.9 & 6.1 &  247.8981 & +37.6048 \\
  BoRG0y & 27.0 & 6.5 &  177.9196 & +54.6847 \\
  BoRG0y & 26.6 & 7.9 &  177.9726& +54.6995 \\
  BoRG0y & 27.2 & 6.3 &  177.9751& +54.6979 \\
  BoRG0j & 26.9 & 5.1 & 178.1887 & +0.9340 \\
  BoRG0c & 26.7 & 6.6 & 118.9794 & +30.7178 \\
  BoRG0g & 26.5 & 6.1 & 124.8104 & +49.1775 \\
  BoRG0t & 26.7 & 8.6 & 117.7142 &+29.2715 \\
  BoRG0t & 27.1 & 6.5 & 117.7064& +29.2977 \\
  BoRG0t & 27.0 & 5.9 & 117.6965& +29.2851 \\
\enddata
\tablecomments{First column: survey field ID. Second column: total
 magnitude in the J band, including aperture correction (automag). Third column: 
detection $S/N$ in the J band (isophotal measurement). Fourth and fifth columns: $Y_{098}$-dropout coordinates (Degrees, J2000 system).}
\end{deluxetable}  

\newpage
\begin{deluxetable}{lrrrrrrr}
\small
\tablecolumns{8}
\tablecaption{$Y_{098}$-dropout ($z\sim8$) sources in field BoRG58. \label{tab:drops}}
\tablehead{\colhead{ID} &{\small $\mathrm{Mag_{tot}}$} & \multicolumn{2}{c}{{\small{Position}}}
&\multicolumn{4}{c}{{\small{ Photometry [isomag]}}} \\
\colhead{} & \colhead{J$_{125}$ {\tiny{[automag]}}} &\colhead{R.A.} &\colhead {Decl.} &\colhead{{\small $V_{606}$}} &\colhead{{\small $Y_{098}$}} &
\colhead {{\small $J_{125}$}} & \colhead{{\small $H_{160}$}}}
\small
\startdata
{\tiny BoRG58\_17871420} & $25.8\pm 0.1$ & 219.2107 &+50.7260  & $>28.6$ & $>28.8$ &$25.9~[13.0\sigma]$& $26.0~[8.0\sigma] $ \\ 
{\tiny BoRG58\_14061418} &  $27.2\pm 0.3$& 219.2241 &+50.7260 & $>29.0$ & $>29.4$ & $27.5~[5.1\sigma]$ & $27.8~[2.6\sigma]$ \\
{\tiny BoRG58\_12071332} & $26.9\pm 0.2$  & 219.2311& +50.7241  & $>28.3$ & $>29.3$ & $27.3~[5.5\sigma]$ & $27.6~[2.7\sigma]$\\ 
{\tiny BoRG58\_15140953} & $27.2\pm 0.2$ &219.2203 &+50.7156   & $>28.5$ &$>29.2$ &  $27.4~[5.4\sigma]$ &$27.7~[2.7\sigma]$ \\
{\tiny BoRG58\_14550613} & $27.0\pm 0.2$ &219.2224 &+50.7081 & $>29.0$ & $>29.3$ & $27.2~[6.2\sigma]$ & $27.6~[2.9\sigma]$ \\
\enddata
\tablecomments{First column: source ID. Second column: total magnitude
(J$_{125}$, including aperture correction (automag). Third/fourth
columns: R.A./Decl. (Deg), J2000 system.  Final columns: source
photometry within detection pixels (isomag), with detection
significance within brackets, or $1\sigma$ upper limits.}
\end{deluxetable}


\end{document}